\documentclass[reprint,amsmath,amssymb,aps,preprintnumbers,nofootinbib,showpacs]{revtex4-1}

\usepackage{amssymb}
\usepackage{graphicx}
\usepackage{subfigure}
\usepackage{fullpage}
\usepackage{colortbl}
\usepackage{hyperref}

\begin{document}

\preprint{UCI-TR-2016-04}

\title{Boosting low-mass hadronic resonances}

\begin{abstract}
Searches for new hadronic resonances typically focus on high-mass spectra,  due to overwhelming QCD backgrounds and detector trigger rates.  We present a study of searches for relatively low-mass hadronic resonances at the LHC in the case that the resonance is boosted by recoiling against a well-measured high-$p_{\textrm{T}}$ probe such as a muon, photon or jet. The hadronic decay of the resonance is then reconstructed either as a single large-radius jet or as a resolved pair of standard narrow-radius jets, balanced in transverse momentum to the probe. We show that the existing 2015 LHC dataset of $pp$ collisions with $\int\mathcal{L}dt = 4\ \mathrm{fb}^{-1}$ should already have powerful sensitivity to a generic $Z'$ model which couples only to quarks, for $Z'$ masses ranging from 20-500 GeV/c$^2$.
\end{abstract}

\author{Chase Shimmin}
\affiliation{Department of Physics and Astronomy, UC Irvine, Irvine, CA 92627}
\author{Daniel Whiteson}
\affiliation{Department of Physics and Astronomy, UC Irvine, Irvine, CA 92627}

\date{\today}

\maketitle

 \section{Introduction}

Searches for resonance peaks in the two-jet invariant mass spectrum are a central feature of the physics program of every collider experiment of the past half-century. Theoretically, this is well motivated due to the many classes of models of new physics which predict $s$-channel resonances with significant couplings to quarks and gluons. Experimentally, the search is attractive as it can be done in a fairly model-independent manner and because increases in center-of-mass energy can provide new sensitivity even in small initial datasets.  

The upper range of sensitivity in terms of the hypothetical resonance mass is limited by the center of mass energy and the fraction of that energy contained in the interacting partons.  In the LHC era, the power to discover or exclude such hadronic resonances has been extended into the TeV range, though no evidence of statistically significant excesses have been seen.    The lower range of sensitivity is controlled by more mundane factors, such as the enormous background rates, which would swamp the trigger and data-acquisition systems.  These high rates demand minimum $p_{\textrm{T}}$ thresholds for the jets which create a lower bound on the sensitivity at a mass of approximately $M=2p_{\textrm{T}}$.  As a result, recent searches have no sensitivity below several hundred GeV, and no experiment has probed below $M=300$~GeV using the dijet final state in the past two decades.  Indeed, in terms of the coupling between quarks and the heavy resonance, limits in this low-mass region are weaker than limits in higher-mass regions~\cite{Dobrescu:2013coa}.

\begin{figure}[h!]
\begin{center}
\includegraphics[width=0.5\columnwidth]{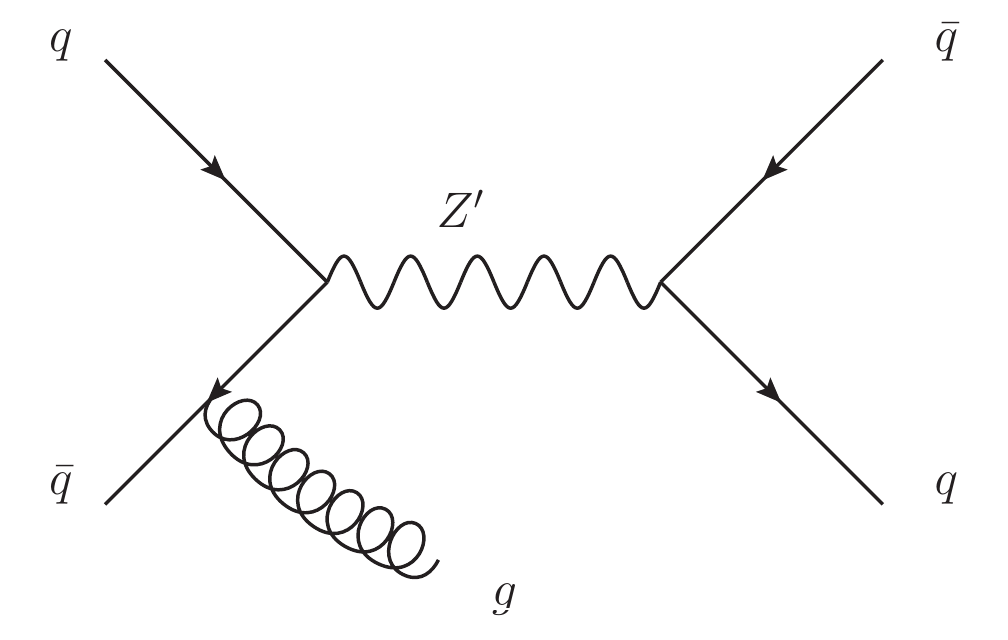}
\includegraphics[width=0.5\columnwidth]{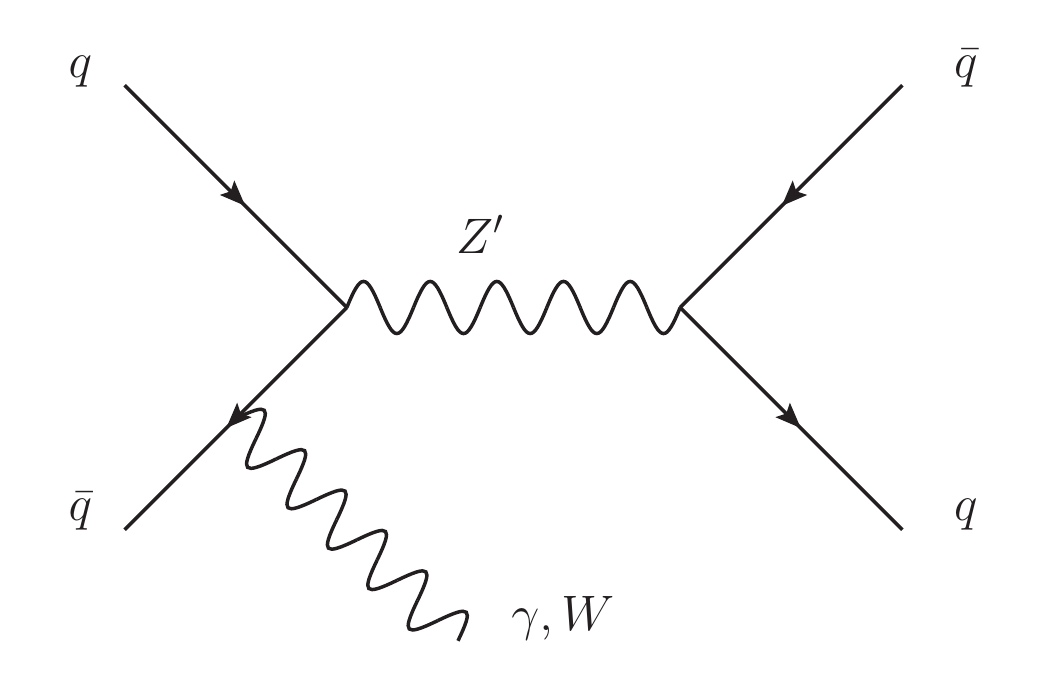}
\caption{Diagrams of $Z'$ production with recoil against either a gluon (top), a photon or $W$ boson (botton).
}
\label{fig:feyn}
\end{center}
\end{figure}

In this paper, we investigate an alternative approach in which the trigger thresholds are avoided by examining data where the light resonance (denoted $Z'$ without loss of generality) is boosted in the transverse direction via recoil from initial-state radiation (ISR) of a photon ($\gamma+Z'$), a $W$ boson ($W+Z'$), or a jet ($j+Z'$); see Fig.~\ref{fig:feyn}.
Requiring a hard ISR object in the final state comes at the cost of  reduced $Z'$ production rates, but allows highly-efficient triggering at much lower $Z'$ masses than typically possible when triggering directly on the $Z'$ decay products.
A similar idea was explored in Ref.~\cite{An:2012ue} in the context of dark matter searches; that study projected limits on low-resonances in all boson ISR channels, and suggests that jet channel is the most sensitive.
Our study builds on this work by accounting for (and ameliorating) the important impact of additional $pp$ interactions ({\it pile-up}), employing more realistic background models\footnote{Specifically, correct handling of jet multiplicity and combinatorics dilutes the power of the $Z'+j$ channel in relation to the $Z'+\gamma$ and $Z'+W$ channels.} as well as considering the power of modern jet substructure tools. 

We show that sensitivity approaching unit couplings can be achieved in the low-mass (20-500 GeV) region using the existing LHC dataset.

\section{The model}

Many models of new physics~\cite{PhysRevD.34.1530,Langacker:2008yv,Salvioni:2009mt} include a new $U(1)$ gauge sector and a new vector boson $Z'$.  Models with a low-mass $Z'$, of the type considered here, are especially appealing as a potential mediator between the standard model and the dark sector~\cite{An:2012va,Rajaraman:2011wf,Goodman:2010ku,Goodman:2010yf}.   Such models can avoid flavor constraints if the couplings to quarks are the same for each generation, and can have significantly larger couplings to quarks than leptons while preserving anomaly cancellations~\cite{Dobrescu:2013coa}.  For the purposes of this study, we consider a simple extension to the standard model with a single extra $Z'$ boson which couples exclusively and equally to all quarks by adding the Lagrangian term:

\begin{equation}
\mathcal{L} \supset \frac{g_B}{6} \bar q \gamma^\mu q Z'_\mu
\end{equation}

\begin{figure}[h!]
\begin{center}
\includegraphics[width=0.7\columnwidth]{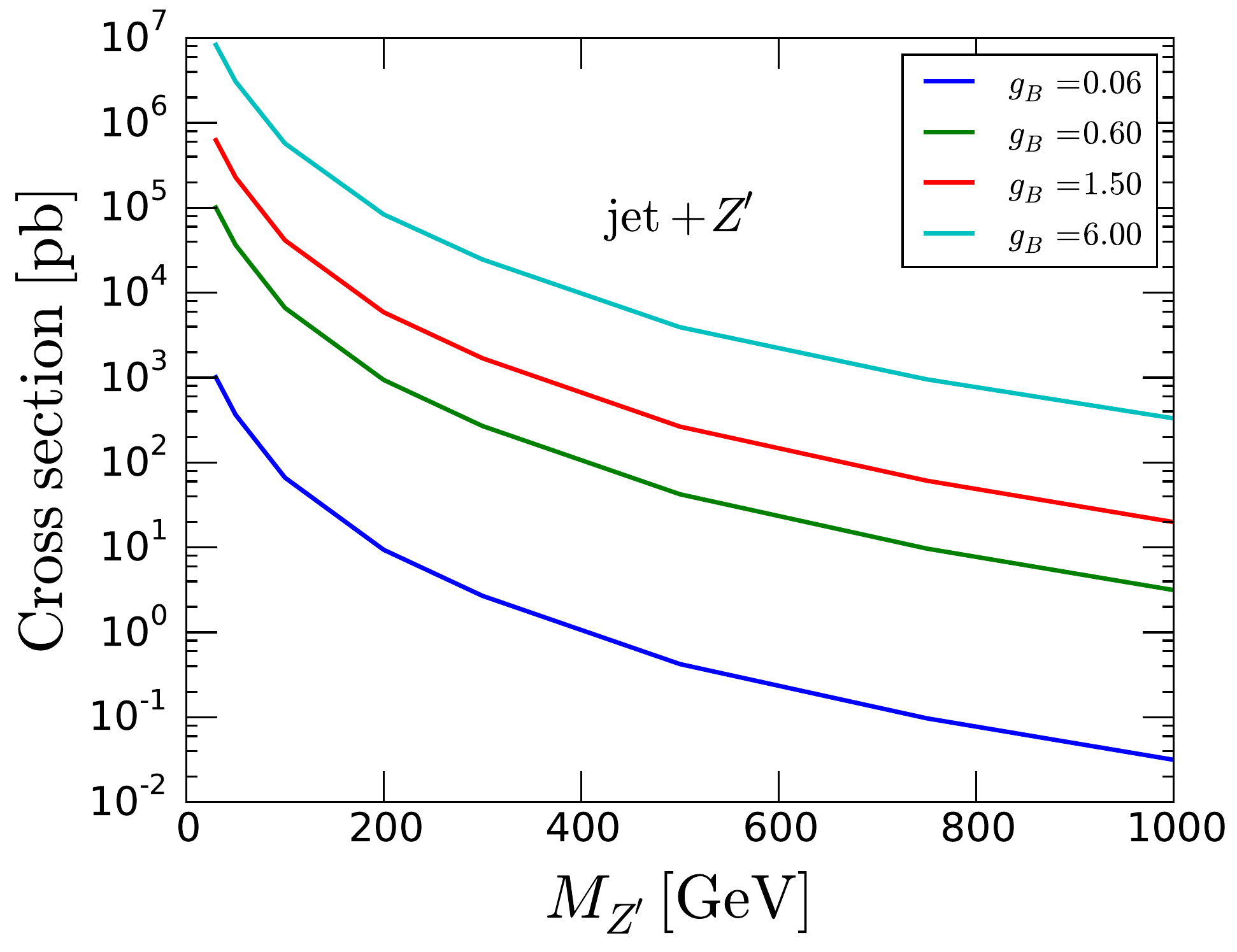}
\includegraphics[width=0.7\columnwidth]{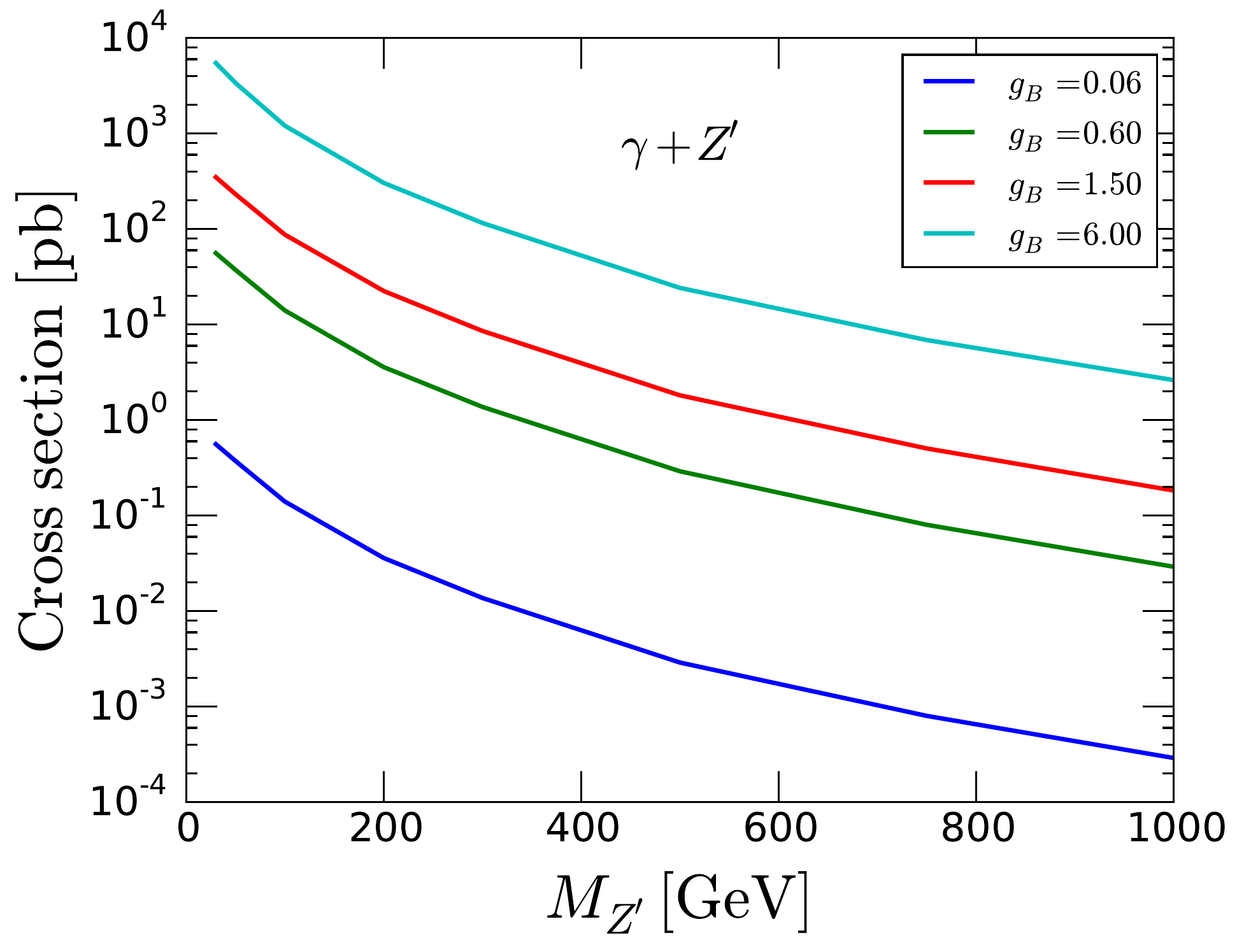}
\includegraphics[width=0.7\columnwidth]{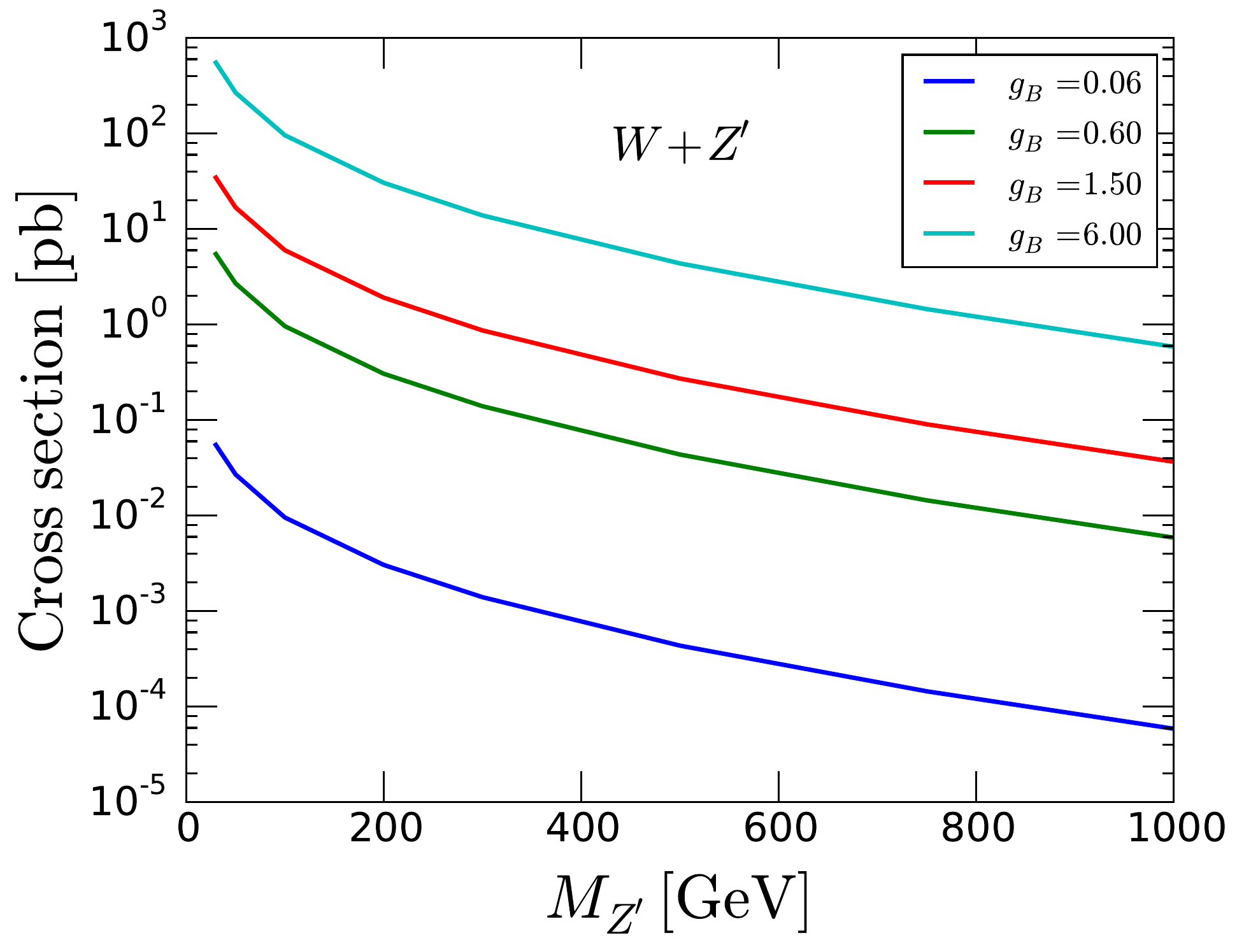}
\includegraphics[width=0.7\columnwidth]{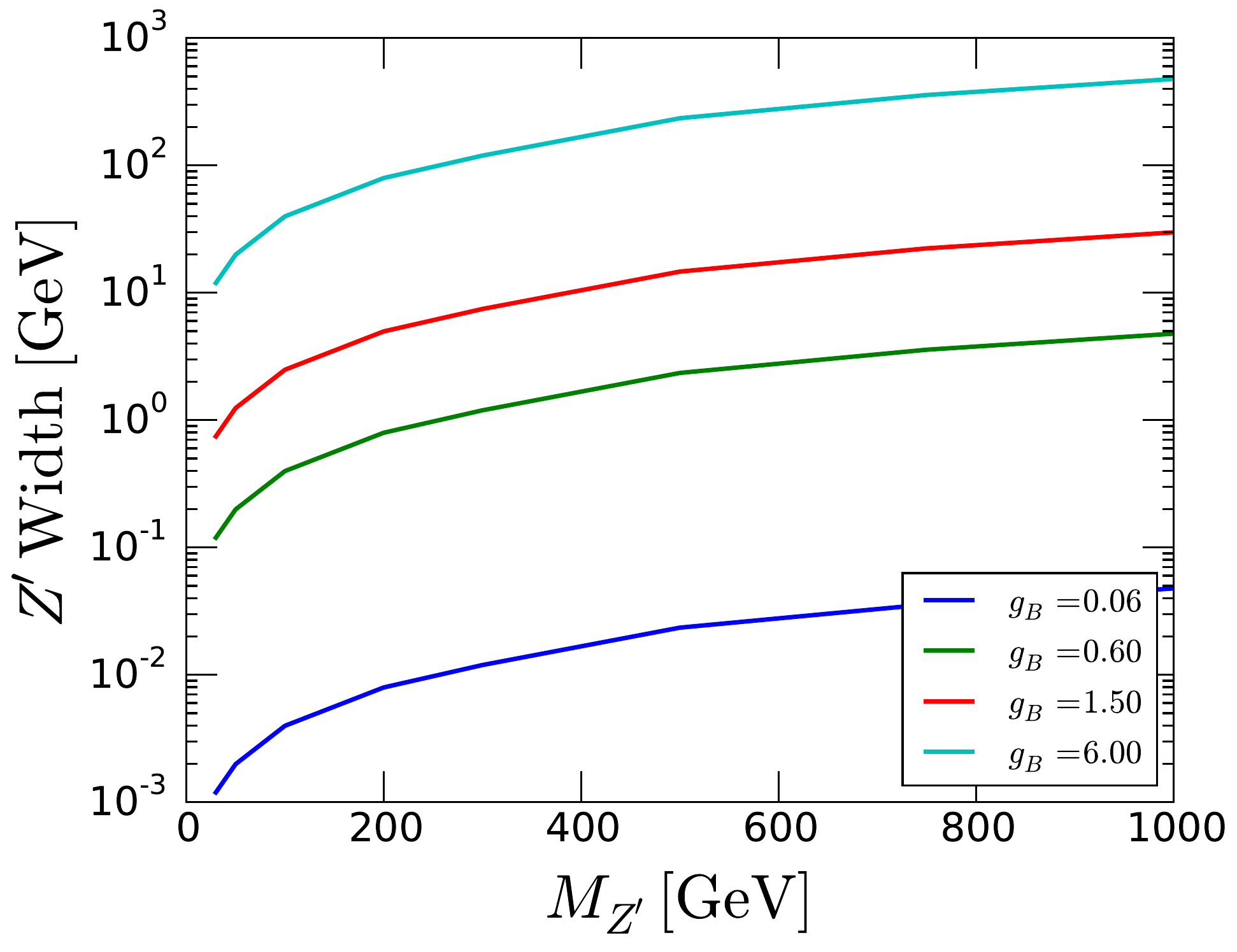}
\caption{Production cross sections at $\sqrt{s}=13$ TeV in $pp$ collisions for $j+Z'$, $W+Z'$  and $\gamma+Z'$.  Also shown is the width of the $Z'$ as a function of mass for several choices of gauge coupling $g_B$.
All calculations include the parton-level requirement of  a jet, charged lepton, or photon in their respective channels with $p_\textrm{T}>10$\ GeV.
}
\label{fig:xs}
\end{center}
\end{figure}

The free parameters of the model are the boson mass, $M_{Z'}$, and the quark coupling constant, $g_B$.
This term arises for example from a gauged baryon number scenario where all quarks have a $g_B/3$ charge under a new $U(1)_B$ field, as described in Ref.~\cite{Dobrescu:2013coa}.
Projected limits on $g_B$ can indicate the feasibility of searching for perturbative theories with features similar to the simplified model here.   Cross sections and widths are shown in Fig.~\ref{fig:xs}.

In these studies, simulated signal samples are generated with $g_B=1.5$. Limits on the cross section can then be converted into upper limits on the gauge coupling strength using the cross section scaling relation, which is approximately $\sigma \propto g_B^4$.
At this value of $g_B$, the $Z'$ width is small comparable to the mass resolution.

\section{Simulated samples}
\label{sec:simulated-samples}

Simulated samples are used to model the kinematics of the signal and background processes.  

Events with a hypothetical $Z'$ boson are are simulated at parton level with {\sc madgraph}5~\cite{Alwall:2011uj}, with {\sc pythia}~\cite{Sjostrand:2006za} for showering and hadronization and {\sc delphes}~\cite{deFavereau:2013fsa} with the ATLAS-style configuration for detector simulation.

The $\gamma+$jets background is generated with {\sc sherpa}~\cite{Gleisberg:2008ta} requiring one photon and $1-3$ additional hard partons.  The multi-jet background is also generated with {\sc sherpa}, requiring $2-4$ hard partons in the final state.

The measurement of jet masses is sensitive to the presence of additional in-time $pp$ interactions, referred to as {\it pile-up} events.  We overlay such interactions in the simulation chain, with an average number of interactions per event of $\left<\mu\right>=15$, which is comparable to the level observed in ATLAS 2015 data with the LHC delivering collisions at a 25ns bunch crossing interval.

The impact of pile-up events on jet reconstruction can be mitigated using several techniques.
First, we employ a jet-area-based pileup subtraction on narrow-radius jets as implemented by {\sc delphes}.  Additionally, when reconstructing large-radius jets, we apply a jet-trimming algorithm~\cite{Krohn:2009th} which is designed to remove pileup while preserving the two-pronged jet substructure characteristic of boson decay.

\section{$\gamma+Z'$ Channel}
\label{sec:gammaZ}

\subsection{Event selection and Reconstruction}
\label{subsec:gammaZ_eventsel}

The photon channel benefits from both the availability of relatively low-$p_{\textrm{T}}$ unprescaled triggers, as well as reduced combinatorial ambiguity in the topology of the final state compared to the jet channel.

For all events in the $\gamma + Z'$ channel, we require at least one isolated photon with $p_{\textrm{T}}^{\gamma} > 120$ GeV, which reflects the threshold of the lowest unprescaled single-photon trigger available to ATLAS in 2015 data.
For signal masses of 300 GeV and above, we additionally require a leading photon with $p_{\textrm{T}}^{\gamma} > 170\mathrm{\ GeV}$, as this provides a slight increase in sensitivity.

The key discriminating feature between the signal and background model is the presence of a resonant peak from the $Z'\rightarrow q\bar q$ decay.
In order to reconstruct the resonance, we examine two techniques.
The first is to simply construct the invariant mass distribution of a pair of standard jets, clustered using the anti-$k_{\textrm{T}}$ algorithm with distance parameter $R=0.4$. We consider all \textit{pileup-subtracted} anti-$k_{\textrm{T}}$ $R=0.4$ jets with $p_{\textrm{T}}^j>20\mathrm{\ GeV}$ and select the pair with the highest $p_{\textrm{T}}$ of the jet-jet system. We refer to this as the {\it dijet} mode below.

As the angular separation of the quarks may be quite small in the case of a very light or very high-$p_{\textrm{T}}$ $Z'$, we consider a second approach of reconstructing a single large-radius jet with distance parameter $R=1.0$. We refer to this as the {\it large-$R$ jet} mode below.  In this mode, we require at least one trimmed anti-$k_{\textrm{T}}$ jet with $R=1.0$ and $p_{\textrm{T}} > 80\mathrm{\ GeV}$ and jet mass of at least $ 20\mathrm{\ GeV}$.
These jets are trimmed by reclustering into $k_{\textrm{T}}$ subjets with $R_{\textrm{trim}}=0.2$ and dropping subjets with less than 3\% of the original jet $p_{\textrm{T}}$.
In the case of multiple large-$R$ jets, the one with greatest $p_{\textrm{T}}$ is selected.

Due to conservation of momentum, the $p_{\textrm{T}}$ of the photon and $Z'$ candidate should be balanced in the final state.
However, due to finite detector resolution effects, soft radiation, and pileup, the reconstructed balance is imperfect.
Hence, we apply the loose requirement that $| p_{\textrm{T}}^{Z'} - p_{\textrm{T}}^{\gamma} | / p_{\textrm{T}}^{\gamma} < 0.5$.
This slightly improves sensitivity by rejecting higher-multiplicity background events where the jet(s) selected do not fully balance the photon, while also improving the signal shape by rejecting events where the wrong jets were selected for reconstruction.

\subsection{ Backgrounds }

The dominant background is due to standard model prompt photon production, labeled $\gamma+$jet throughout. Sherpa has been shown~\cite{ATLAS-PUB-2015-016} to accurately model events with photons and jets in various kinematic distributions.  No $k$-factor is available in the literature, so in the results below we demonstrate the effect of a $k$-factor ranging from 1 to 2.

We also account for standard model $\gamma+W$ and $\gamma+Z$ production; simulated samples are generated at leading order in $\alpha$ with {\sc madgraph}5; note that these processes are approximately three orders of magnitude below the $\gamma$+jet background, and approximately one order of magnitude below the predicted  rate for the hypothesized $Z'$ signal with $g_B=1.5$.

\begin{figure}[h!]
\begin{center}
\includegraphics[width=0.9\columnwidth]{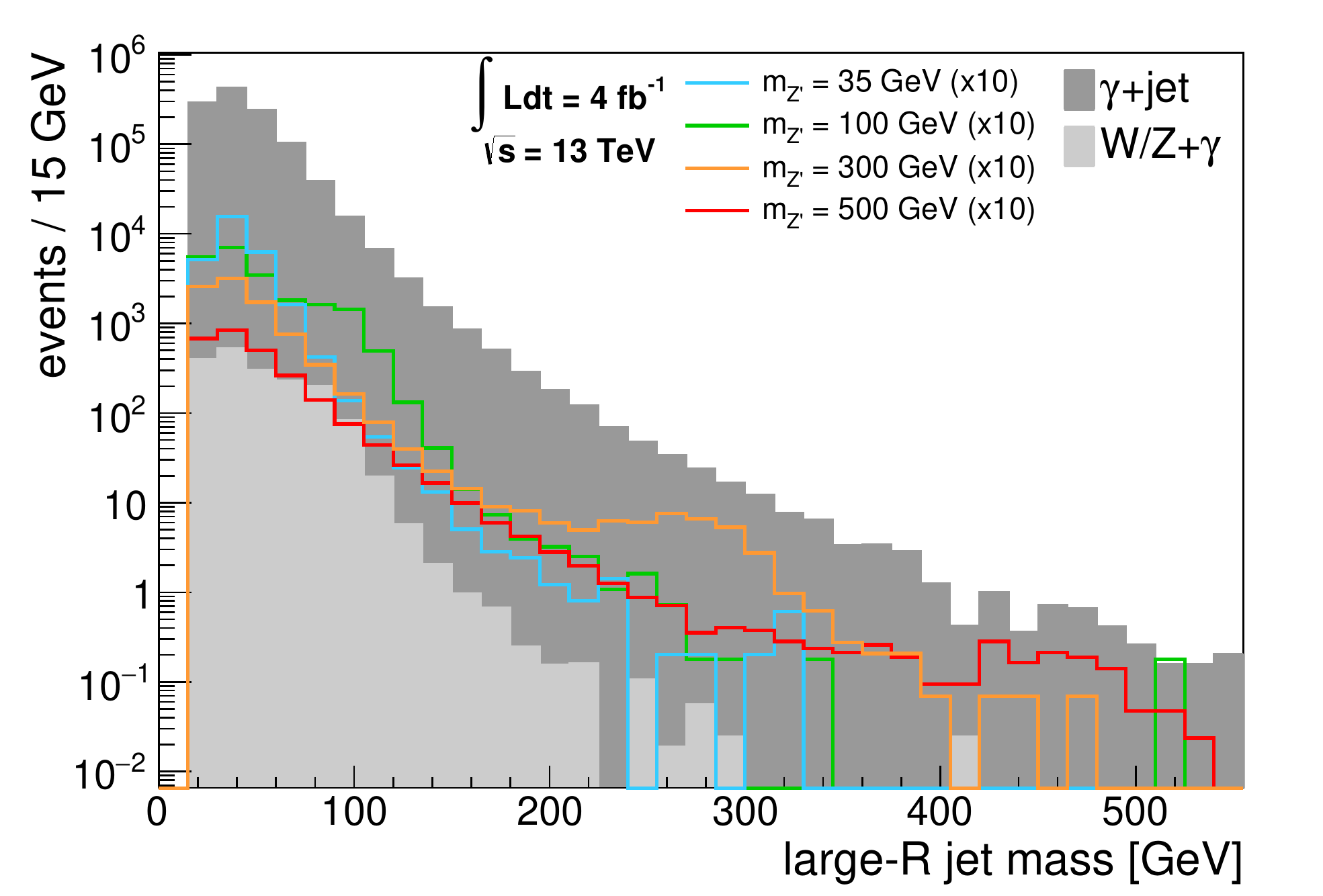}
\includegraphics[width=0.9\columnwidth]{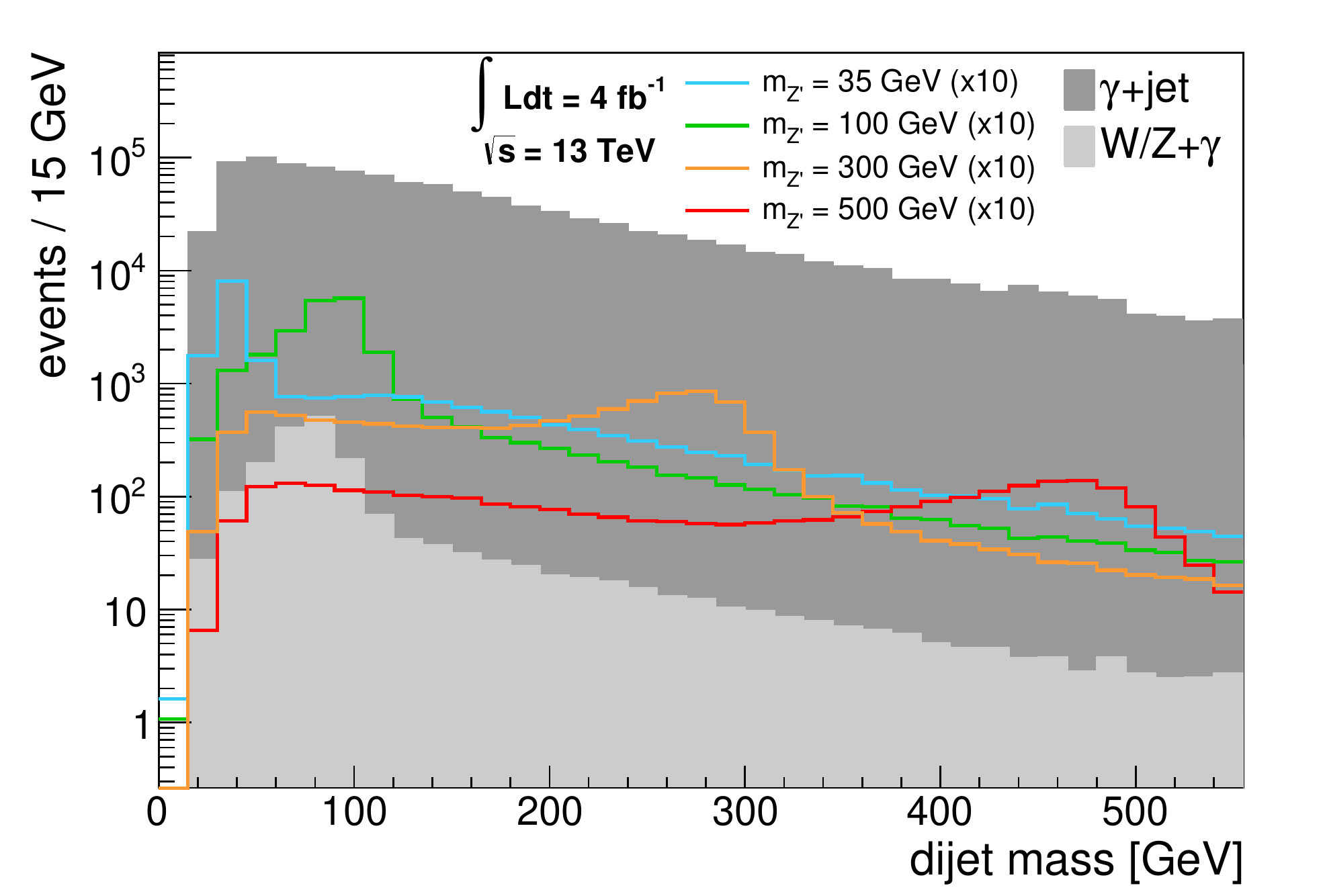}
\caption{
Distribution at $\sqrt{s}=13$ TeV and $\int \mathcal{L}dt = 4\mathrm{fb}^{-1}$ of reconstructed $Z'$ candidate mass in the $\gamma+Z'$ channel, for both the single large-$R$ jet (top) and the resolved dijet case (bottom) considered in the text.
Also shown are signal distributions, generated with $g_B=1.5$, and scaled by a factor of 10 for visibility.
}
\label{fig:yj}
\end{center}
\end{figure}

Figure~\ref{fig:yj} shows the distribution of reconstructed large-$R$ or dijet masses in both signal and background processes for the $\gamma+Z'$ channel.

\section{$W(\mu\nu)+Z'$ Channel}
\subsection{Event selection and Reconstruction}

As with the photon channel, the leptonic $W$ channel has little ambiguity in selecting the final state jets, and benefits from the lower-$p_\textrm{T}$ lepton trigger levels, which potentially enhances resolution of very light resonances by limiting the collimation of decay products of the less-boosted object.
This comes at the cost of lower branching fractions of both the vector boson ISR and leptonic decay mode, which greatly reduces the signal production cross section.
For simplicity, we consider only the muon final state; adding electrons increases the complexity of the analysis and at best results in a factor of $\sim \sqrt{2}$ in cross section sensitivity, translating to only approximately $9\%$ improvement in $g_B$ reach.

For all events in the $W(\mu\nu) + Z'$ channel, we require exactly one isolated muon with $p_{\textrm{T}}>40\ \mathrm{GeV}$, representing the muon trigger.
Events containing additional electrons or muons with $p_\textrm{T}>10\ \mathrm{GeV}$ are vetoed.

To select the $Z'$ candidate in the large-$R$ and dijet cases, the same procedure as described in Sec.~\ref{sec:gammaZ} is followed.
However, since the observed $\mu$ alone is not expected to balance the $p_\textrm{T}$ of the resonance, no momentum conservation cut is applied.

\subsection{ Backgrounds }

In contrast to the $\gamma+Z'$ channel, backgrounds to the $W+Z'$ channel are not wholly dominated by a single process.

The largest source of background is due to standard model $W$ boson production with additional ISR jets, with the $W$ decaying leptonically, referred to as $W+$jets throughout.
This background is generated using {\sc sherpa} by sampling events with a final state containing $\mu+\nu_\mu$ and up to 2 additional partons; a parton-level requirement the invariant mass $m(\mu \nu_\mu)$ is greater than 2 GeV is imposed.

We also account for backgrounds due to  SM top single- and pair-production, $Z+$jets with leptonic decays, and semileptonic diboson processes; each of these is generated with {\sc madgraph}5.
The $Z+$jets background is somewhat reduced by the additional lepton veto; however, due to the relatively low-$p_\textrm{T}$ muon threshold, many events contain soft additional leptons which are not reconstructed, and hence pass the selection.
The sole background to show resonant structure in the reconstructed jet mass is the diboson $WZ$ production with semileptonic decay.

\begin{figure}[h!]
\begin{center}
\includegraphics[width=0.9\columnwidth]{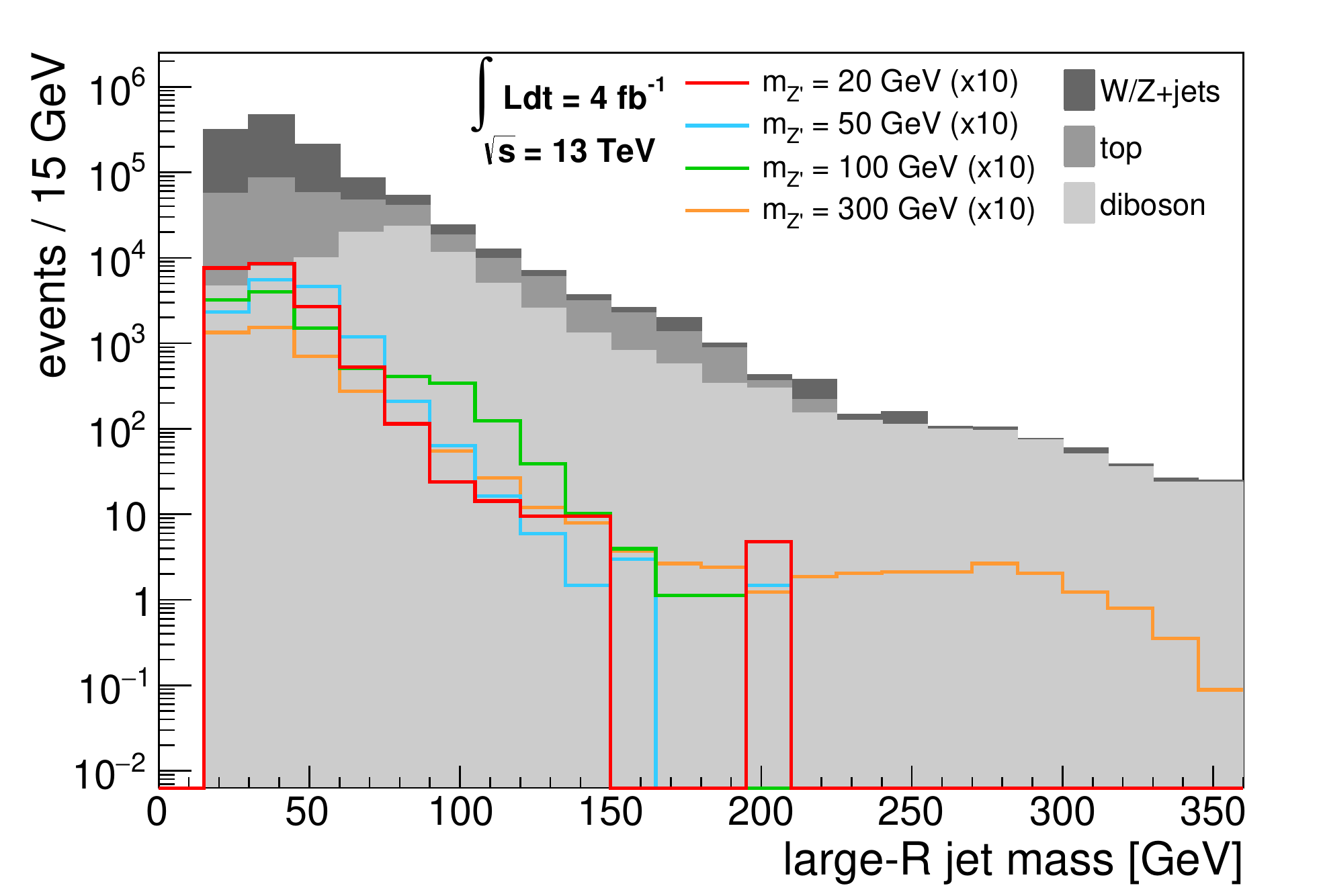}
\includegraphics[width=0.9\columnwidth]{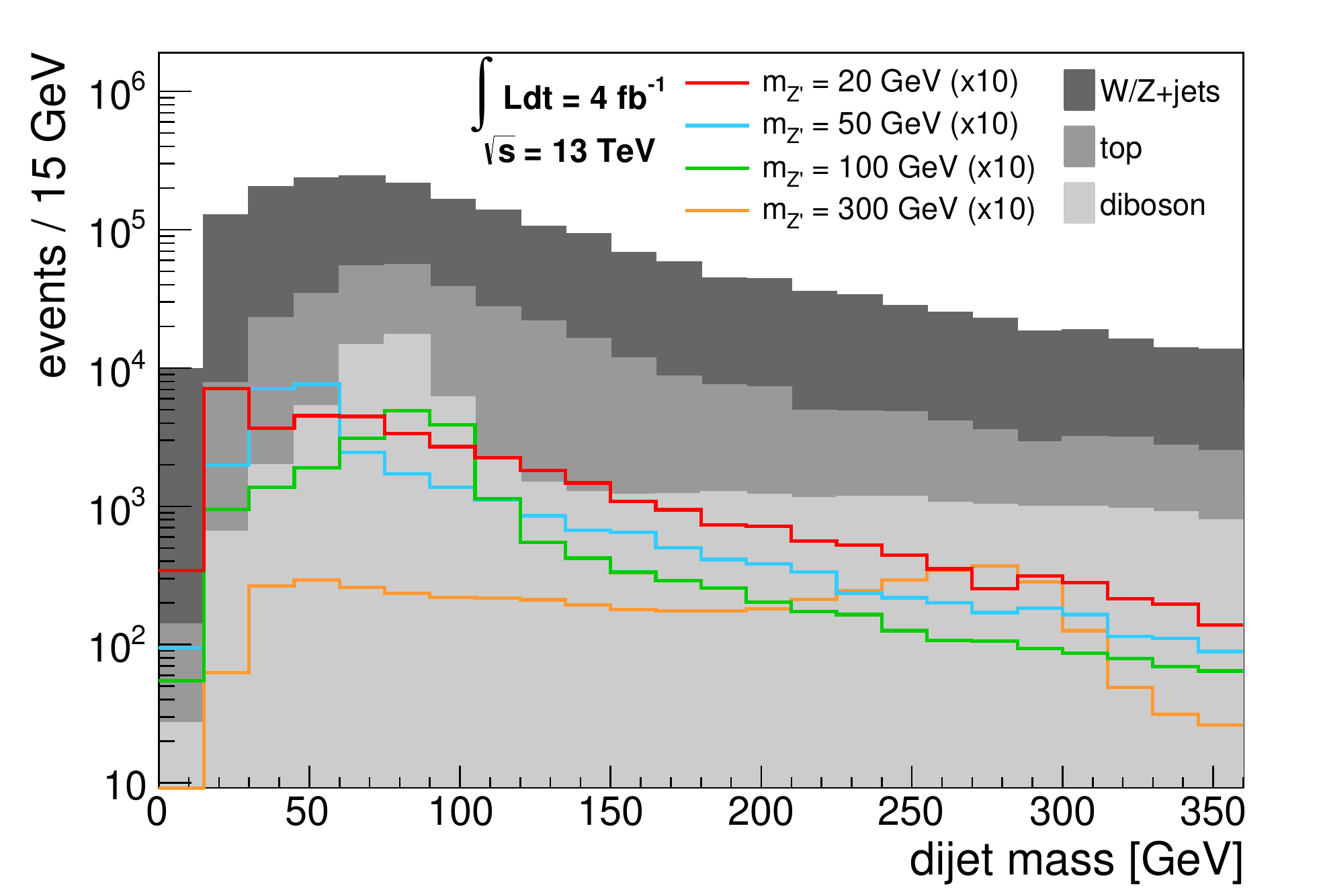}
\caption{
Distribution at $\sqrt{s}=13$ TeV and $\int \mathcal{L}dt = 4\mathrm{fb}^{-1}$ of reconstructed $Z'$ candidate mass in the $W(\mu\nu)+Z'$ channel, for both the single large-$R$ jet (top) and the resolved dijet case (bottom) considered in the text.
Also shown are signal distributions, generated with $g_B=1.5$, and scaled by a factor of 10 for visibility.
}
\label{fig:wj}
\end{center}
\end{figure}

Figure~\ref{fig:wj} shows the distribution of reconstructed large-$R$ or dijet masses in both signal and background processes for the $W+Z'$ channel.

\section{Jet $+Z'$ Channel}

\subsection{ Event selection}

The jet$+Z'$ channel contains only jets in the final state, leading to greater ambiguity in defining the reconstructed $Z'$ mass.
Here we will refer to the reconstructed \textit{decay jet(s)} as either a single large-$R$ jet or a pair of resolved jets which define the reconstructed mass of the hypothetical $Z'$ decay. We refer to the \textit{probe jet} as the small-$R$ jet which is opposite in momentum to the decay jet(s).
It is impossible to always assign the decay jets correctly, particularly in the presence of additional QCD radiation and pileup.
While the simple heuristic approaches described below work reasonably well, further studies may benefit considerably from the use of multivariate techniques in order to select the most signal-like jet(s) from each event.

For all events in the jet$+ Z'$ channel, we require at least one anti-$k_\textrm{T}$ $R=0.4$ jet satisfying $|\eta| < 3.2$ and $p_{\textrm{T}} > 360$ GeV; this represents the lowest unprescaled single-jet triggers available to ATLAS in 2015 data in the central detector.

For the large-$R$ jet reconstruction, we use the same anti-$k_{\textrm{T}}$ jets with $R=1.0$ and trimming as described in Sec.~\ref{sec:gammaZ}.
To avoid the possibility of selecting a probe jet overlapping with a candidate large-$R$ jet, we examine all pairs of reconstructed $R=0.4$ and $R=1.0$ jets, and consider only those pairs which are separated with a $\Delta R > 0.8$.
We select the pair with highest large-$R$ jet $p_{\textrm{T}}$.
In cases where this is not unique, we then choose the jet with highest small-$R$ jet $p_{\textrm{T}}$.
The small-$R$ jet is assigned as the probe jet, while the large-$R$ jet is taken as the $Z'$ candidate.

For the dijet case, the $Z'$ candidate is built from the pair of small-$R$ jets whose combined four-momentum has the largest $p_{\textrm{T}}$.
Of the remaining unassigned jets, the small-$R$ jet with largest $p_{\textrm{T}}$ is assigned as the probe jet.

As before, in order to require momentum balance in the underlying event, we require that the $Z'$ candidate satisfy $| p_{\textrm{T}}^{Z'} - p_{\textrm{T}}^{\textrm{probe}} | / p_{\textrm{T}}^{\textrm{probe}} < 0.5$. 

\subsection{ Backgrounds }

The overwhelming background is standard model QCD multi-jet production, and is modeled using {\sc sherpa} as described in Sec. \ref{sec:simulated-samples}. The large rate of this background requires in a high single-jet $p_\textrm{T}$ threshold of $360\textrm{\ GeV}$ at ATLAS and $450\textrm{\ GeV}$ at CMS.

\begin{figure}[h!]
\begin{center}
\includegraphics[width=0.9\columnwidth]{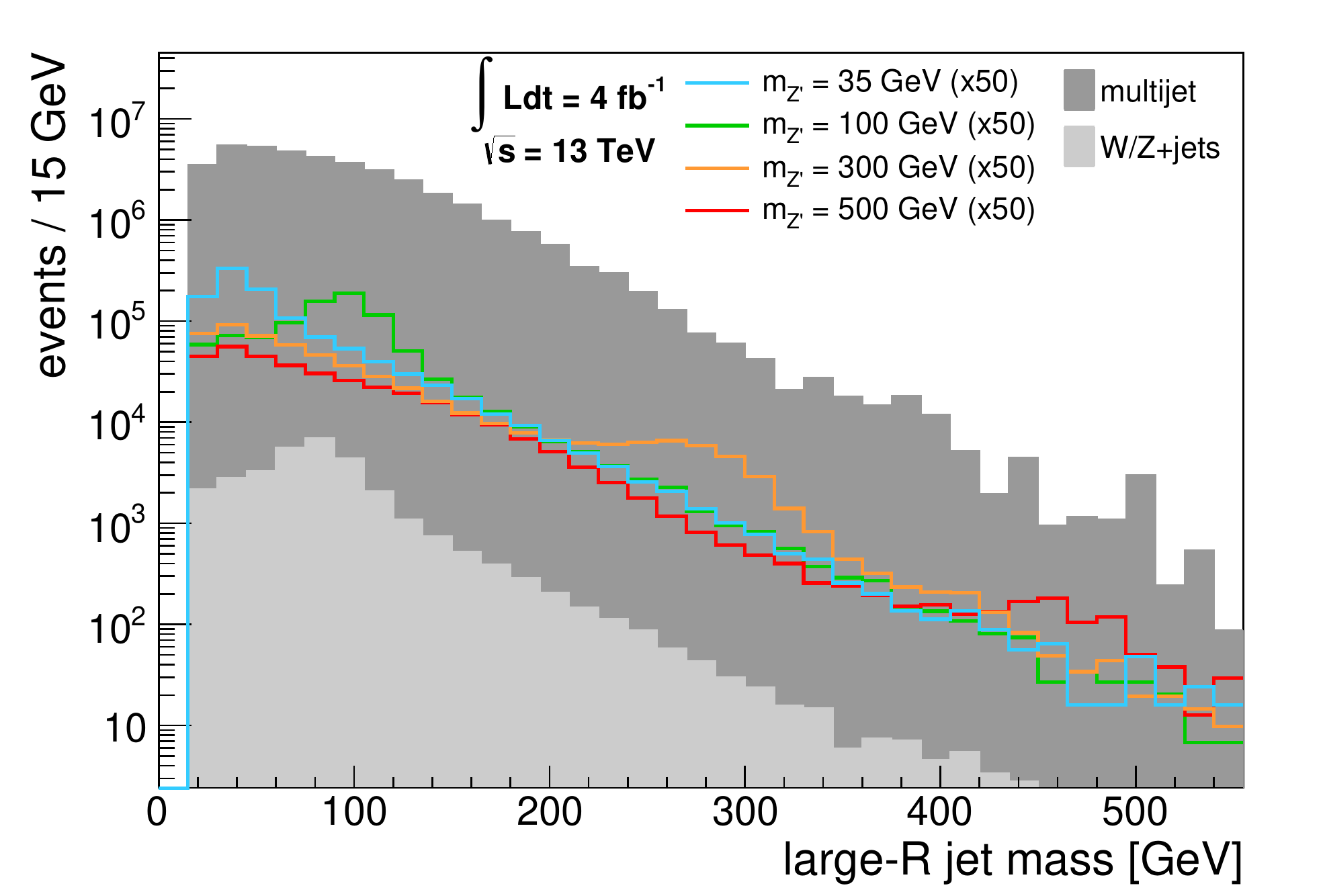}
\includegraphics[width=0.9\columnwidth]{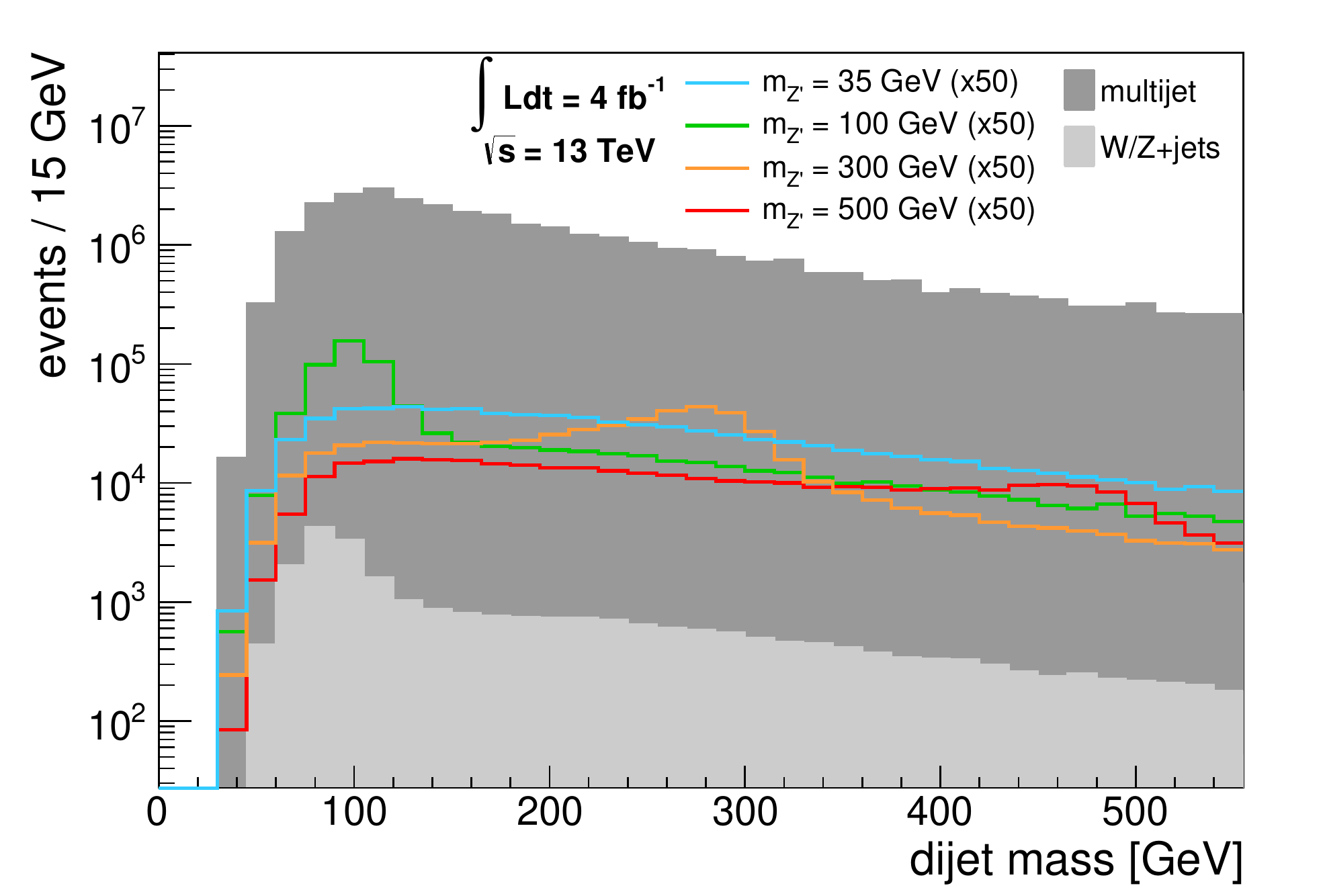}
\caption{
Distribution at $\sqrt{s}=13$ TeV and $\int\mathcal{L}dt=4\mathrm{fb}^{-1}$ of reconstructed $Z'$ candidate mass in the jet$+Z'$ channel, for both the single large-$R$ jet (top) and the resolved dijet case (bottom) considered in the text.
Also shown are signal distributions, generated with $g_B=1.5$, and scaled by a factor 50 for visibility.
}
\label{fig:jj}
\end{center}
\end{figure}

We also account for standard model $W$ and $Z$ boson production, in association with one hard parton, using MadGraph. However, as in the $\gamma+Z'$ channel, the contributions are very small relative to the other backgrounds. Figure~\ref{fig:jj} shows the distribution of reconstructed $Z'$ candidate masses in both the large-$R$ or dijet cases for  signal and background processes for the jet+$Z'$ channel.

\section{ Sensitivity }

The estimates of the signal and background yields for $pp$ collisions corresponding to $4\textrm{ fb}^{-1}$ of luminosity are used to calculated expected upper limits on the production of the hypothetical $Z'$ signal.
Limits on the cross section are then converted into upper limits on the gauge coupling strength $g_B$ between quarks and the $Z'$.

Limits are calculated at 95\% CL using a profile likelihood ratio~\cite{Cowan:2010js} with the CLs technique~\cite{Read:2002hq,Junk:1999kv} and a binned distribution in the reconstructed mass of the hypothetical $Z'$ boson.   In the $\gamma+Z'$ and $W+Z'$ channels, the large-$R$ jet mass distribution is binned from $24 - 550\textrm{\ GeV}$, with bin widths increasing from $8 - 50\textrm{\ GeV}$; the dijet mass distribution is binned every 10 GeV from $10-520\textrm{\ GeV}$.
In the jet$+Z'$ channel, the large-$R$ jet mass distribution is binned from $24 - 620\textrm{\ GeV}$ with bin widths increasing from $8 - 50\textrm{\ GeV}$, and the dijet mass is binned every 10 GeV from $80-550$ GeV.  Hypothetical signal masses are considered only if the reconstructed mass contains a localized peak, allowing for normalization of the background and profiling of the nuisance parameters in sidebands. The resolved dijet case requires a minimal angle between the decay products and hence a minimal reconstructed mass, suppressing a resonance peak below 50 GeV, as seen in Fig.~\ref{fig:jj}; the large-$R$ jet case does not suffer from this issue.

Several sources of systematic uncertainties on the signal and background processes are considered. The dominant $\gamma$+jet and multi-jet backgrounds are assigned a 15\% uncertainty on the overall normalization.\footnote{ Accurate estimates of these uncertainties would come from studies of the scale dependence of the $k$-factor; such information is not available in the literature. We choose 15\% and 5\%, for the QCD and EW background respectively,  typical values for such uncertainties.  Note that due to the use of a profile likelihood technique, these uncertainties can be significantly constrained in data using background-dominated sideband regions above and below the hypothetical $Z'$ mass. Therefore the resulting statistical limits are not very sensitive to the initial assignment of the systematic uncertainty.} The smaller $W/Z+X$ backgrounds are assigned 5\% uncertainties. More significant may be the uncertainty in the expected $Z'$ reconstructed mass distribution, especially for values of $M_{Z'}$ which give reconstructed distributions that are more difficult to distinguish from the background distributions. A significant source of uncertainty in the reconstructed mass distribution may come from the calibration of the hadronic response and the overall jet calibration.  Detailed studies from experimental collaborations are needed for definitive statements, but to approximate the impact of such sources of uncertainty, we shift the response of all calorimeter towers by $\pm5\%$.

\begin{figure}[ht!]
\begin{center}
\includegraphics[width=0.95\columnwidth]{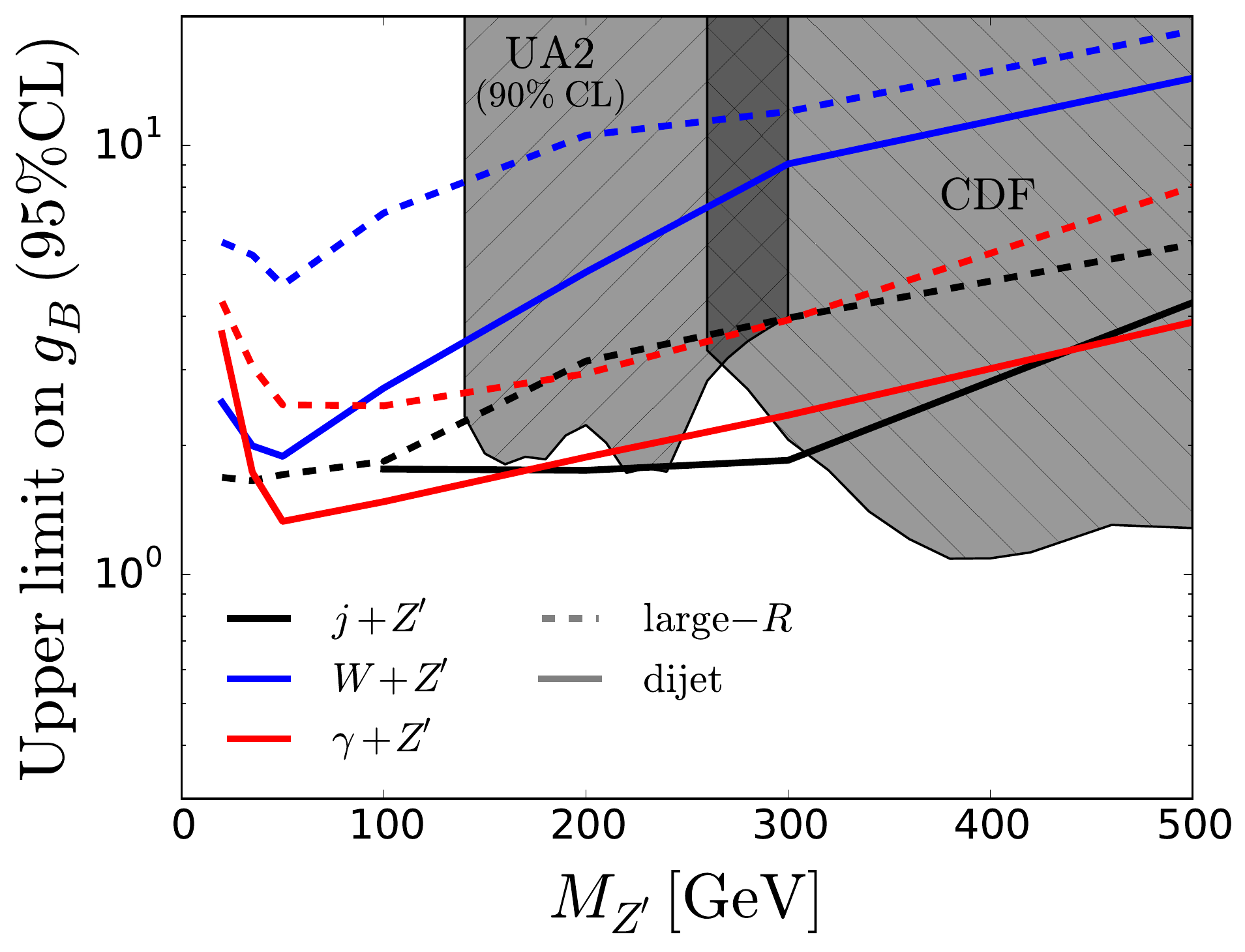}
\caption{ Expected upper limits at 95\% CL on the coupling $g_B$ between the hypothetical $Z'$ boson and quarks, for the $\gamma+Z'$, $W+Z'$, and jet$+Z'$ channels, in both the single large-$R$ jet as well as the dijet modes, for values of $M_{Z'}$ from 20-500 GeV. If the $k$-factors for the largest backgrounds are doubled, the limits are weakened by 17-21\%. The limits are calculated for the case of $pp$ collisions at $\sqrt{s}=13\mathrm{\ TeV}$ with $\int\mathcal{L}dt = 4\ \mathrm{fb}^{-1}$. For comparison, we include existing limits from UA2 and CDF (shaded contours), as interpreted by~\cite{Dobrescu:2013coa}.
}
\label{fig:limits}
\end{center}
\end{figure}

The expected 95\% confidence level limits are shown in Tab. \ref{tab:limits}, and Fig. \ref{fig:limits}.
We note that while the resolved dijet technique tends to perform better at all but the lowest masses considered in our study, the application of further jet substructure techniques may prove to enhance the sensitivity of the large-$R$ jet method.

\begin{table}[h]
\centering
\begin{tabular}{r l ccccccc}
\hline \hline
 & & \multicolumn{7}{c}{ $M_{Z'}$\ [GeV] } \\
 Channel \vspace{1.5mm} & mode & 20 & 35 & 50 & 100 & 200 & 300 & 500 \\
 \hline 
  \vspace{0.5mm}$\gamma+Z'$  & dijet &3.6 &	1.7 &	1.3 &	1.5 &	1.9 &	2.4 &	3.9\\
&large-$R$ jet & 4.3 &	3.1 &	2.5 &	2.5 &	2.9 &	3.9 &	8.0\\
 \hline 
  \vspace{0.5mm}$W+Z'$  & dijet &2.5 &	2.0 &	1.9 &	2.7 &	5.1 &	9.1 &	14.3\\
&large-$R$ jet & 6.0 &	5.6 &	4.7 &	7.0 &	10.6 &	12.0 &	18.5 \\
 \hline
 \vspace{0.5mm}jet$+Z'$ & dijet & 1.7 &	1.7 &	1.7 &	1.8 &	3.1 &	4.0 &	5.9 \\
&large-$R$ jet & -- &	-- &	-- &	1.8 &	1.7 &	1.8 &	4.3 \\
 \hline \hline
\end{tabular}
\caption{  Expected upper limits at 95\% CL on the coupling $g_B$ between the hypothetical $Z'$ boson and quarks, for the $\gamma+Z'$, $W+Z'$, and jet$+Z'$ channels, in both the single large-$R$ jet as well as the resolved dijet modes, for values of $M_{Z'}$ from 20-500 GeV.
The limits are calculated for the case of $pp$ collisions at $\sqrt{s}=13\mathrm{\ TeV}$ with $\int\mathcal{L}dt = 4\ \mathrm{fb}^{-1}$.
}
\label{tab:limits}
\end{table}

\section{Discussion}

We have presented the expected experimental sensitivity for hadronic resonances, in a low-mass region typically inaccessable using traditional search strategies. By requiring that the $Z'$ recoil against a jet, a $W$ boson or a photon, we are able to escape the high-threshold trigger requirements and suppress the background.

Generally, limits using the dijet reconstruction method are stronger but tend to lose sensitivity when the angle between the decay quarks is small enough that the $R=0.4$ jets merge.
This happens at approximately $M_{Z'} \sim 0.4\ p_\textrm{T} / 2$, where the $p_\textrm{T}$ scale is determined by the trigger threshold in each channel.
Once the jets merge, the dijet and large-$R$ reconstruction methods perform comparably, although both eventually degrade in sensitivity at lower masses as the low-mass portion of the background becomes difficult to fit.

The large-$R$ jet method is not as sensitive to collinear decay products, and so therefore should be more robust at lower masses. Improvements in large-$R$ jet reconstruction techniques will possibly allow experiments to set limits at masses even lower than those projected here.

Although we studied the use of substructure variables such as $N$-subjettiness~\cite{Thaler:2010tr} and Energy Correlation Functions~\cite{Larkoski:2013eya} for large-$R$ jets, we found these variables did not lead to a selection with reliably improved sensitivity. It is possible that with the more sophisticated detector modeling available to experimentalists, including more realistic tracking, vertexing, and calorimeter clustering, improved pileup removal and substructure resolution may enhance limits in this channel, particularly at very low masses.

Compared to the results of Ref.~\cite{An:2012ue}, we find that once pileup and detector effects are accounted for, the photon channel has sensitivity much more comparable to the jet channel, with reach to even lower $Z'$ masses.
Although these effects also considerably reduce the overall expected sensitivity, we show that sensitivity approaching unit couplings can be achieved in the low-mass (20-500 GeV) region using the existing LHC dataset.

Several important challenges remain for an experimental analysis, most notably the construction of a reliable background estimate that can be constrained in data.
The limits presented here assume the possibility of constraining the background model by fitting the mass sidebands simultaneously with the signal hypotheses.
In practice, this would most likely be accomplished using a parametric fit function; this approach would be easiest to validate in the region of the mass spectrum which is monotonic and smooth, possibly limiting the reach towards the lowest-mass resonances.
If experimentalists can develop methods to overcome these challenges, the potential for discovery exists with data available today.

\section{Acknowledgements}

The authors would like to acknowledge Mohammad Abdullah, Linda Carpenter, Sam Meehan, Felix Yu, and Ning Zhou for helpful discussion and comments. We also thank Anthony DiFranzo for providing the MadGraph implementation of the $Z'$ model.

\bibliography{zplite}

\end{document}